\documentclass{emulateapj}

\usepackage{graphicx}
\usepackage{epstopdf}
\newcommand{\myemail}{silas@gemini.edu}

\shorttitle{CXO J005428.9-723107}
\shortauthors{Laycock \& Drake}

\begin{document}


\title{A Flaring X-ray Source with an H$\alpha$-bright Counterpart toward the SMC.}


\author{Silas Laycock}
\affil{Gemini Observatory, 670 N. A'ohoku Pl. Hilo, HI, 96720,  \myemail}

\author{Jeremy J. Drake}
\affil{Harvard-Smithsonian Center for Astrophysics, 60 Garden St, Cambridge, MA, 02138}



\begin{abstract} 
We report the discovery of a flaring X-ray source with an optical counterpart with H$\alpha$ emission and red-excess, in the direction of the SMC.
A 100 ksec X-ray observation with Chandra detected a flare lasting $\sim$6 ksec in the source CXO J005428.9-723107. The X-ray spectrum during the flare was consistent with a thermal plasma of temperature kT=2.5$\pm$0.4 keV. In quiescence following the flare the spectrum was softer (kT$\sim$ 0.4 keV). Timing analysis did not reveal any significant periodicities or QPOs.
Optical images taken with the Magellan-Baade 6.5m telescope show a single star in the (0.9") error circle. This star has apparent magnitude V=19.17, exhibits enhanced H$\alpha$ emission (H$\alpha$--r = -0.88$\pm$0.02), and has a large proper motion.  Alternative explanations are explored, leading to identification as a relatively nearby (Galactic) coronally active star of the BY Draconis class.
\end {abstract}


\section{Introduction}
The Chandra SMC Deep Fields project \citep{laycock2008} has observed two regions in the Small Magellanic Cloud to a flux limit of 10$^{-15}$erg/cm$^2$/s. The fields were selected as the most rich in high mass X-ray binary (HMXB) pulsars (Deep-Field-1/ DF1), and the location of the youngest stellar population (DF2). The primary aim is to locate the unseen majority of HMXBs, by detecting them in the quiescent state between outbursts. The deep observations complement  ongoing weekly monitoring with {\it RXTE} that has now monitored the outburst activity of pulsars in the SMC for 8 years \citep{galache2008}. Other targets of great interest for the {\it Chandra} program are LMXBs, binaries containing black holes, and Cataclysmic Variables (CVs), none of which have yet been found in the SMC. 

At the faint fluxes reached in these observations, a large by-catch of background active galactic nuclei and foreground (galactic) sources are also present in the data. The SMC lies at high galactic latitude (b=-60$\degr$) so foreground contamination is sparse.  The principal source of foreground stars is the galactic thick-disk and halo populations (see e.g \cite{bahcall1984}). The most probable foreground objects to be detected in X-rays are stellar coronae, including active binaries,  and (to a lesser extent) CVs. Active stellar coronae are a feature of solar and subsolar-mass stars.  An era marked by frequent powerful flares 10-100 times the brightness of those seen on the sun is thought to be experienced by most if not all dwarf stars, typically during the early portion of their lives (see \cite{mathu1990}). In this paper we describe our discovery of such a star, which is the most prominent representative of the sample seen in the SMC Deep Fields survey.

\section{X-Ray Discovery of a Flaring Source}
SMC Deep Field-1was observed  observed by {\it Chandra} for 100 ksec in a pair of $\sim$ 50 ksec pointings (same roll angle),  over the course of two days Apr 25-26, 2006. The Advanced CCD Imaging Spectrometer (ACIS) was used, with the ACIS-I array at the aimpoint. The observations and data reduction are described by \cite{laycock2008}.

\begin{figure}
\begin{center}
\includegraphics[angle=-90,width=8cm]{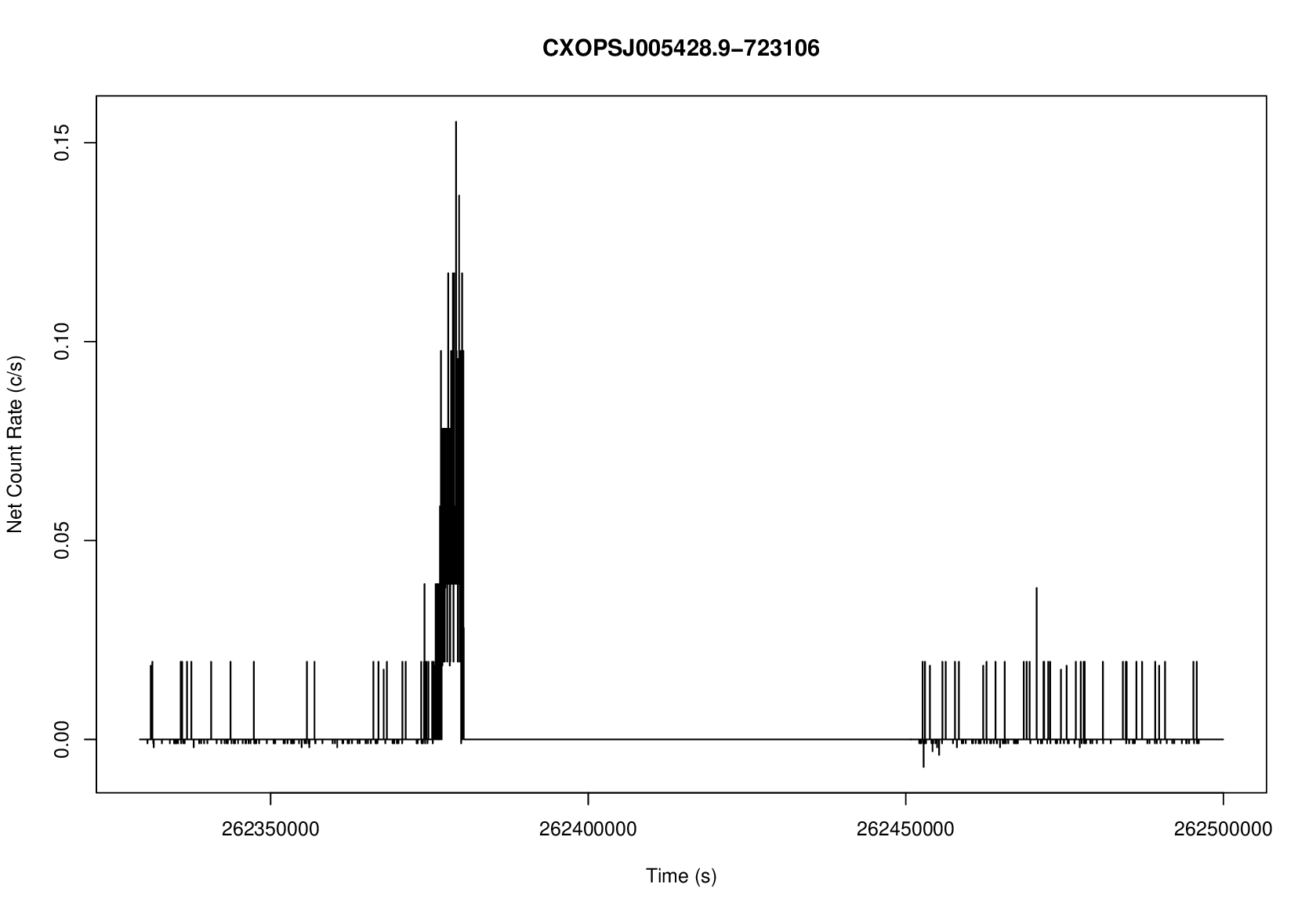}
\includegraphics[angle=-90,width=8cm]{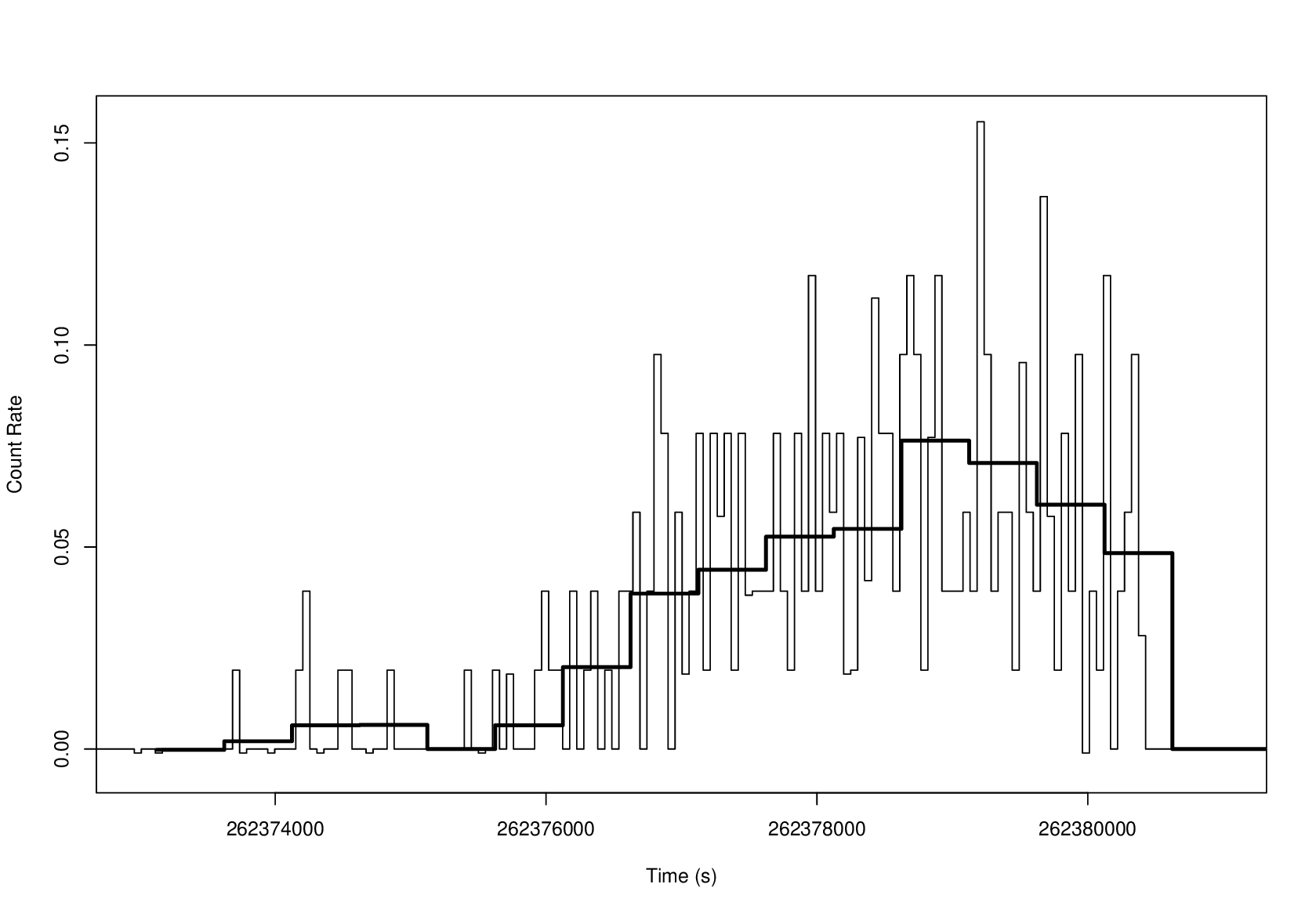}
\caption{{\bf  Lightcurve of the flaring source CXO J005428.9-723107. Left panel is the full lightcurve incorporating both observations, with initial binning at 16 times the 3.2 Chandra ACIS frame interval (51.2s). Right panel is a zoomed-in view of the flare, the thick line shows the lc binned at 500s. }}
\label{flarelc}
\end{center}
\end{figure}

Lightcurves binned by 16 times the 3.2s ACIS frame time were visually inspected for all sources with at least 50 counts in the stacked dataset. This search  turned up a large flare in CXO J005428.9-723107, (RA=00:54:28.91, Dec=-72:31:07.12, 95\% error radius=0.53'', including 0.75'' aspect uncertainty\footnote{http://cxc.harvard.edu/cal/ASPECT}=0.91'') lasting for $\sim$5000 sec at the end of the first of two 50 ksec exposures of Deep Field 1, shown in Figure~\ref{flarelc}. None of the other 95 sources so examined showed any similar feature. Due to the gap in coverage, it is possible that the final part of the decay phase was not observed.

\begin{figure}
\begin{center}
\includegraphics[angle=-90,width=10cm]{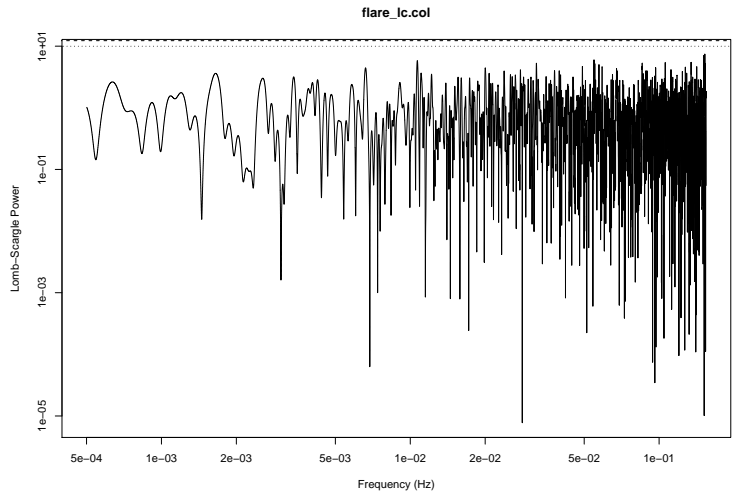}
\caption{{\bf  Power Spectrum of the flare in source CXO J005428.9-723107. The input lightcurve is 7000 sec of unbinned data containing the event. The Lomb-Scargle power-spectrum is plotted in log-log scaling to look for red-noise and QPOs. No identifiable features can be seen, indicating that the source is unlikely to be pulsar or magnetar, unless it has a spin period below the 3.2s sampling interval.}}
\label{flarepds}
\end{center}
\end{figure}

X-ray variability is the hallmark of accretion and is also characteristic of flaring stellar coronae. Cataclysmic variables (accreting WDs) and Low-Mass X-ray binaries (LMXBs: both NS and BH systems) exhibit aperiodic variability (flickering) over a wide range of timescales from milliseconds up. Such variability produces a characteristic power-law distribution (red noise) in the power spectrum. The Lomb-Scargle \citep{scargle} power spectrum of the flare is shown figure~\ref{flarepds}, no periods QPOs or red noise are evident.

\begin{figure}
\begin{center}
\includegraphics[angle=-90,width=10cm]{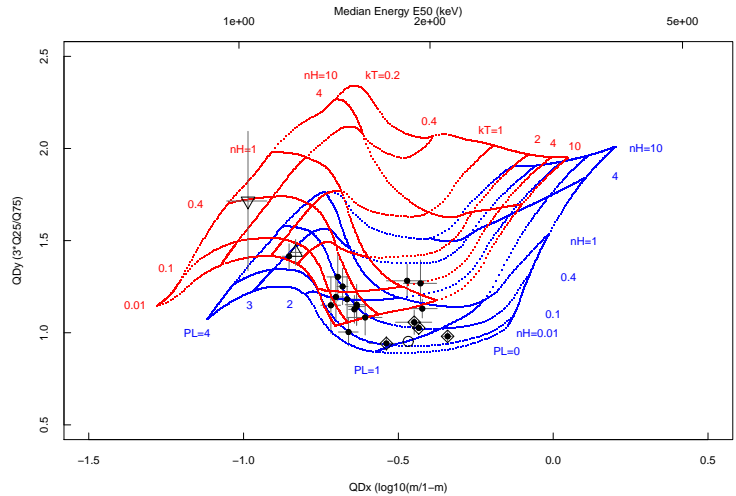}
\caption{{\bf  Quantile Diagram for sources in SMC Deep Field 1 with Signal-to-Noise Ratio $>$10. Two grey triangular points show the flaring source. The upward triangle is the value during the flare, and the downward triangle is after the flare. The black point close to the flaring value is the result for the stacked data. Spectral model grids are shown for power-law plus absorption (Blue) and thermal bremsstrahlung (Red).  Other sources that can be seen are Pulsars and AGN.}}
\label{fig:quantiles}
\end{center}
\end{figure}

\begin{figure}
\begin{center}
\includegraphics[angle=-90,width=10cm]{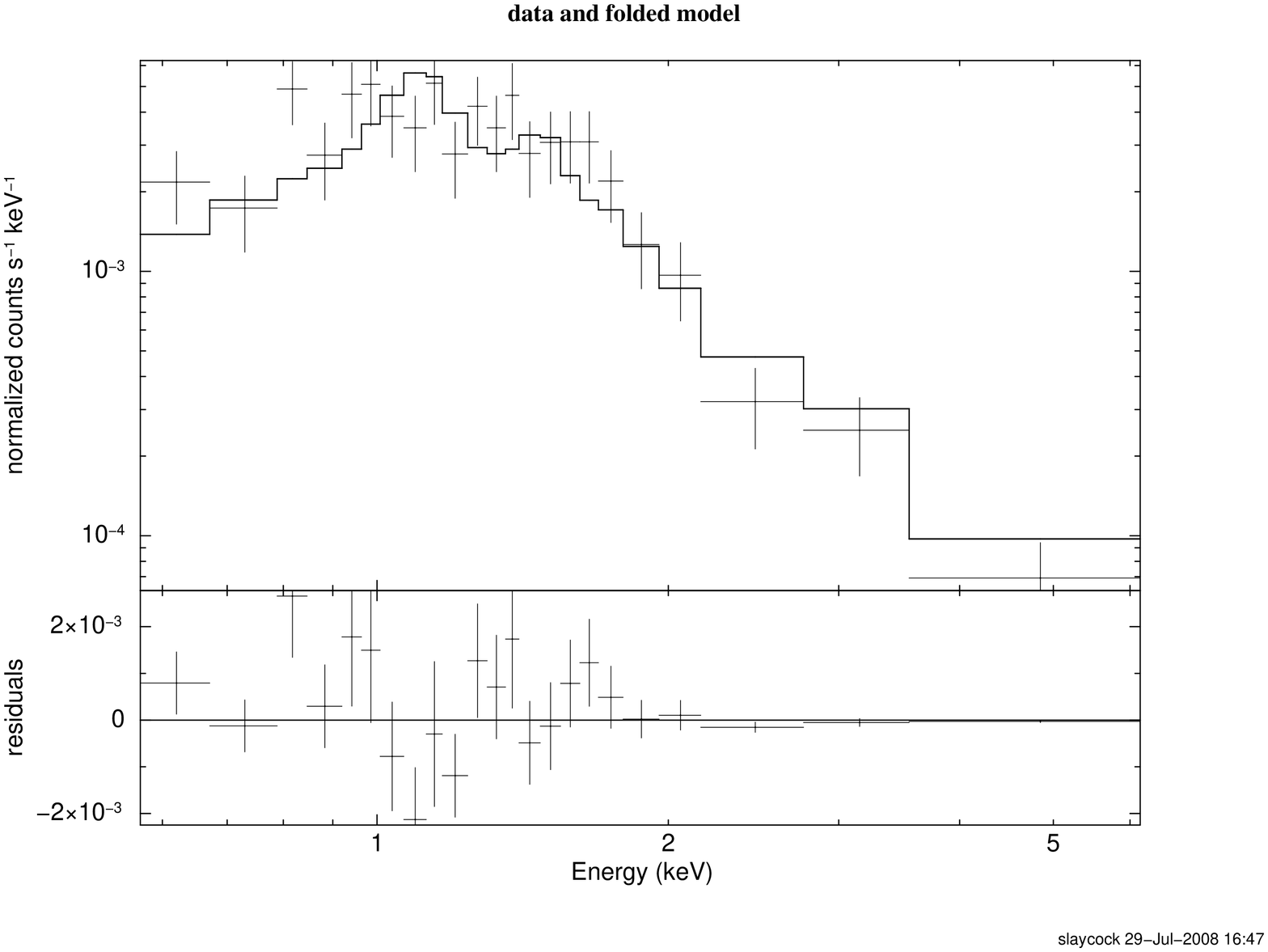}
\caption{{\bf X-ray  Spectrum of  CXO J005428.9-723107 during the flare.  The fitted model is a Raymond-Smith (optically thin thermal plasma) with plasma temperature kT=  2.5 +/-  0.4 keV }}
\label{flarespec}
\end{center}
\end{figure}

Subtracting the total net counts for the source during the second exposure from the first (259cts - 32cts), we obtain an estimate of 277 counts in the flare. We used the quantile technique \citep{hong2004}  to compare the source spectrum between the active and inactive states. The flare (median energy E50 = 1.288 keV) was harder than the quiescent emission (E50 = 1.021 keV).  
The quantile values during and post flare are plotted in  Figure~\ref{fig:quantiles}, along with other bright X-ray point sources for comparison. During quiescence the quantile location is consistent with a fairly cool thermal plasma (kT$\sim$ 0.4 keV) and the extinction is clearly low although poorly constrained. During the flare the quantile moves to higher temperature (kT$\sim$1keV) and we obtain a better constraint on the extinction, confirming its low value (nH$\leq$10$^{21}$ cm). A fit to the X-ray spectrum of the observation containing the flare is shown in figure~\ref{flarespec}. The fitted model is a optically-thin thermal plasma \citep{raymondsmith},  with solar abundance, plasma temperature kT=2.5$\pm$0.4 keV and essentially zero extinction. The reduced $\chi ^2$ = 1.14 (20 degrees of freedom). Fixing the extinction at nH=10$^{21}$ makes the  $\chi ^2$  slightly worse (1.4).  The Mekal model \citep{merwe1985} gave consistent results (kT=1.7$\pm$0.3 keV, $\chi^2$=1.2). For comparison we fit an absorbed power-law ($\Gamma$=3$\pm$0.3, nH=2.6($\pm$1)$\times$10$^{21}$  $\chi ^2$=0.64).  

The peak flux reached in the lightcurve binned by 16 frame intervals was 0.15 counts/sec. The number of counts in the peak bin is 7 or 8 (0.15 c/s = 7.68 counts / 51.2 second bin time.). Applying the best-fit spectral model (RS, 2.5 keV see above) we obtain a peak flux of 1$\times$10$^{-12}$ erg/cm$^2$/s. 

By binning more more coarsely,  a smooth light-curve was obtained at 500 sec per bin. Using this light-curve the flare reaches a peak of 0.0763$\pm$0.0125 counts/s at 3000 seconds from its start. The peak flux in this case is 5.1$\times$10$^{-12}$ erg/cm$^2$/s. During the rising portion of the flare, the count rate climbs at the roughly constant rate of  2.5$\times$10$^{-5}$ count/s/s. The rate of decline after peak is 1.4$\times$10$^{-5}$ c/s/s, slower than the rise-rate by a factor of 0.56. The observations stopped 2 ksec after the peak, before the count rate had dropped to its quiescent level. 

\section{Optical Counterpart}
The field was observed with the IMACS imager mounted on the Magellan-Baade 6.5 meter telescope at Las Campanas, Chile on August 4th, 2007 (07:00 UT).
IMACS was in f/2 configuration providing a circular field of view 30' in diameter, sampled at 0.2''/pixel. Long and short exposures were made through Sloan-g,r,i and Hydrogen-$\alpha$ filters. 

The images were processed using standard IRAF tasks to perform the usual CCD reductions. Astrometric calibration was performed with the 2MASS catalog as a reference. Residuals were of order 0.2" rms  (1 pixel) for TAN (IRAF tangent-plane projection geometry) plate solutions to each CCD frame. The individual frames were reduced independently and combined using the SWARP utility to re-project and stack. All frames were normalized to ADU/second units and combined using weight-maps constructed from flat-fields and saturated pixel masks. This approach combines long and short exposures into a single image.

\begin{figure}
\begin{center}
\includegraphics[angle=0,width=16cm]{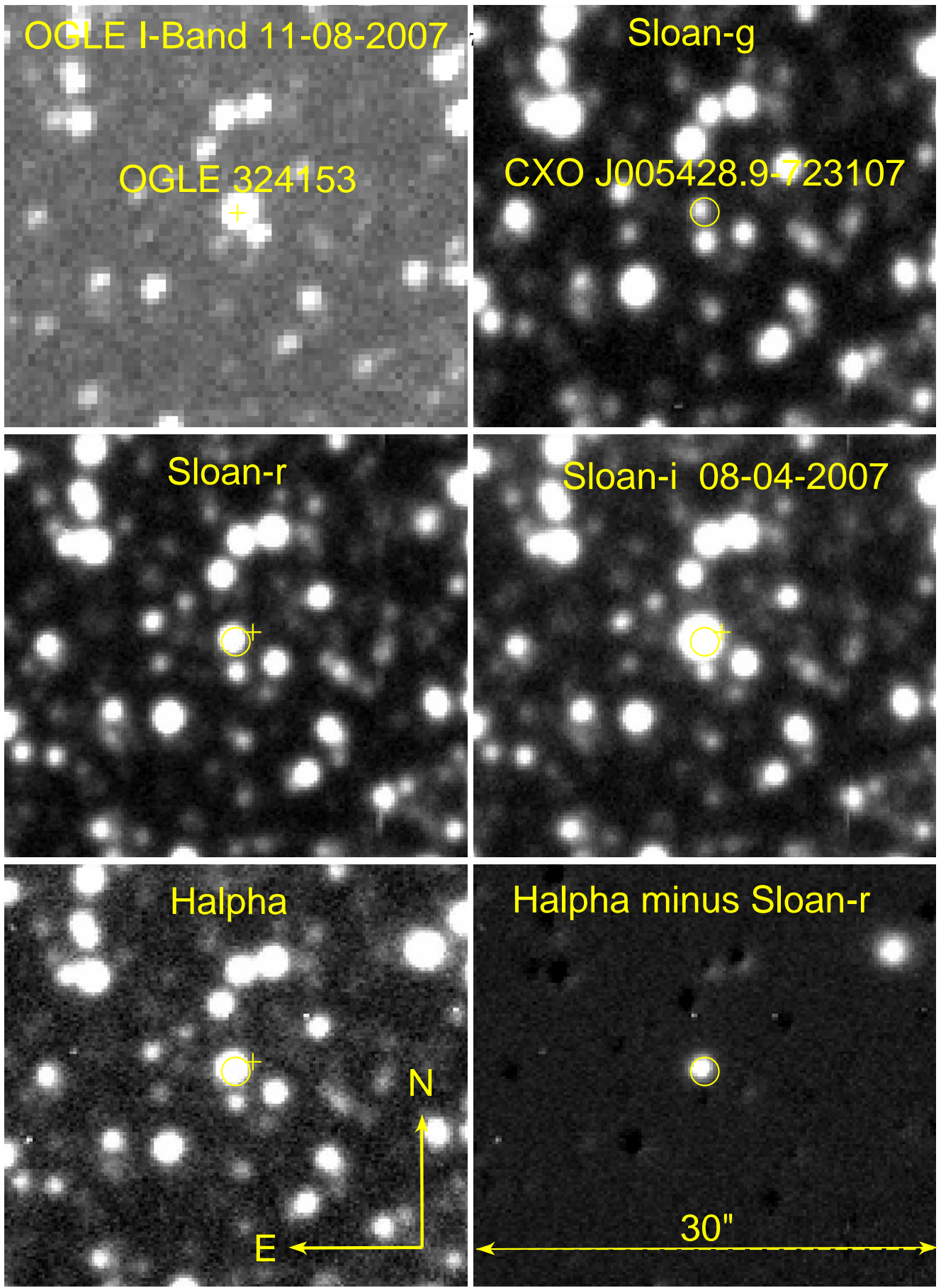}
\caption{{\bf  Optical images of the location of CXO J005428.9-723107. The over-plotted circle indicates the  X-ray position (April 25, 2006) 95\% confidence radius including contributions from ACIS/wavedetect and aspect uncertainty. The star at the center of the displayed 30'' field is the proposed optical counterpart. It is brighter at longer wavelengths and the H$\alpha$-r difference image reveals strong H$\alpha$ emission compared to stars in the field.  Comparison with the OGLE reference image from 1997 shows 1.6'' motion of the star in the intervening decade. The 1997 position is indicated by a cross on the August 2007 IMACS images)}}
\label{fig:optimage}
\end{center}
\end{figure}

For this paper we restricted our analysis to a 1000x1000 pixel sub-region centered on the co-ordinates of  CXO J005428.9-723107. The complete optical survey of the Chandra SMC Deep Fields will be presented by \cite{antinou2009}.  Initial identification of the optical counterpart was performed by visual inspection of the images. A single star was found in the Chandra error circle, as can be seen in Figure~\ref{fig:optimage}. By eye it is evident that the star is substantially redder than most of its neighbors, and is bright in the H$\alpha$ image.

Hydrogen-$\alpha$ (6563$\AA$) in emission is ubiquitous in accretion-powered X-ray sources. Stars with H$\alpha$ line emission can be swiftly identified by the image-subtraction technique, performed on a pair of images taken through a broad-band red filter, and a narrow-band  H$\alpha$ filter.  Examination of our H$\alpha$ and Sloan-r sub-images showed very similar point-spread-functions (PSF), with FWHM$\sim$1'' . An empirical flux scaling factor of 48 was applied to the H$\alpha$ image to compensate for the different filter transmissions. The IMACS narrow band filter is an interference filter which exhibits spatial variations in its bandpass when used in the steeply converging f/2 beam. The effective bandwidth of the H$\alpha$ filter at the location of the sub-image is accordingly calculated to be 1475/48 = 30$\AA$   (bandwidth of the Sloan-r filter is 1475$\AA$). The result of subtracting the flux-scaled H$\alpha$ image from the Sloan-r image is shown in the bottom right panel of Figure~\ref{fig:optimage}.  A bright point-like residual lies in the error circle, while nearly all other stars have cancelled out. The flux in the difference-image residual ($\Delta $F) implies a magnitude difference ($\Delta m$) of -0.882 magnitudes as given by Equation~\ref{eqn:magdiff} 

\begin{equation}
\Delta m = 2.5 log_{10}(F_r/(F_r + \Delta F))
\label{eqn:magdiff}
\end{equation}

where F$_r$ is the background subtracted flux in the Sloan-r image.
The corresponding emission line equivalent width EW(H$\alpha$) $\simeq$ 38$\AA$, assuming a flat continuum and the bandpass derived above.   
 
Photometry of the IMACS sub-field was calibrated against the UBVR catalog of \cite{massey}. The calibration residuals between our magnitudes and the Massey magnitudes transformed to the Sloan system \citep{fukugita1996} were rms$_g$=0.18, rms$_r$=0.11, rms$_i$=0.18.  For the optical counterpart of CXO J005428.9-723107 we measured the following magnitudes: g=19.90$\pm$0.03, r=18.63 $\pm$0.01, i=16.97$\pm$0.01. After transforming back to the standard system, we obtain V=19.17, B-V=1.43, V-R=0.80, R-I=1.70.
The B-V color index of the star suggests spectral type M0, while R-I  supports a later subtype M5+. Large extinction does not work as an explanation of the star's color because its effects would be strongest at shorter wavelengths, exactly opposite to what is observed. In addition, the X-ray spectrum shows there is little to no absorption. A pattern of rising red-excess at longer wavelengths is typical of stars with extended envelopes, an interpretation strongly supported by the extreme H$\alpha$ emission. Alternatively a metal-poor star would partially explain the colors being bluer in B-V than in R-I, compared to a solar metallicity star.
The flux ratio log(fx/fr) is a useful diagnostic because it is independent of distance.  Using Equation~\ref{eqn:fluxratio} we obtained log(fx/fr)[flare]=+0.84 and log(fx/fr) (quiescent)=-1.17, noting the X-ray and optical measurements were not made at the same time.

\begin{equation}
R = log(F_{x}/F_{r})=log(F_{x}) + 5.5 + R/2.5
\label{eqn:fluxratio}
\end{equation}

\section{Discussion}
We have identified an optical counterpart to the Flaring X-ray source CXO J005428.9-723107 in or toward the SMC. The star has apparent V magnitude 19.17, B-V=1.4 and a red excess accompanied by H$\alpha$ in emission. We explore several alternative explanations. Those motivated primarily by the X-ray data include: Symbiotic, Low-mass X-ray binary (LMXB), Cataclysmic variable (CV), Magnetar, and from the optical data: young-stellar object (YSO) or foreground galactic magnetically active dwarf.  Each of these scenarios are consistent with the presence of X-ray flares, plus an optical counterpart with H$\alpha$ emission. 

\subsection{Identification as a Galactic Active-Star or Binary}
Dwarf M stars are the most numerous spectral class in the galaxy, with dMe subtype characterized by H$\alpha$ emission making up about half \citep{mathu1990}. 
Observations of X-ray flaring in the dMe-star Ross 154  described by \cite{wargelin2008}, show striking similarities to the event seen in CXO J005428.9-723107.  They report a flare with a rapid rise to peak in $\sim$1.6 ksec, followed by a slightly slower fading to 10\% of peak in a further 2 ksec, after which there is a very slow decay to quiescence. The quiescent X-ray spectrum of Ross 154 was a thermal plasma with kT=0.46-0.98 keV (depending on model used) rising to 2.95-3.71 keV during the flare. The peak X-ray luminosity was 1.8$\times$10$^{30}$ erg/s.  Ross 154 is a M3.5 dwarf at a distance of just 2.97 pc. The \cite{wargelin2008} spectral fits are nearly identical to the values estimated for CXO J005428.9-723107.

If we adopt the same luminosity as was reported in the Ross 154 flare for CXO J005428.9-723107, then the observed X-ray flux implies a distance of between 94 and 118 parsec, depending on whether the 50 sec or 500 sec binned light-curve is used for the flux estimate. The corresponding distance modulus lies in the range 4.87 to 5.36 mag. The absolute magnitude M$_V$ of the optical counterpart would then be +13.8 to +14.3, This is improbably faint given the observed color (B-V=1.43) indicates a spectral type M0 (M$_V$=+9).  The absolute magnitude can be freely adjusted by raising the distance and flare luminosity as follows: In the absence of significant extinction and assuming an M0V star, $\mu$=10 and D=1kpc, leading to an upper limit on the peak X-ray luminosity of order 10$^{32}$erg/s. Instead assuming M5V, leads to $\mu$=7.2, D=275pc, and a lower limit on L$_X$=4.27$\times$10$^{30}$.  The SMC lies at high galactic latitude, the scale height of disk dwarfs is 350 pc  \citep{bahcall1984}, so a mid-M classification would place the system within the thick-disk component of the galaxy. 

The range in luminosity in the above scenarios can be seen in the most active M-stars, and are frequently attained in BY Draconis stars, which are a species of close binary comprising K and M dwarfs, with magnetic activity excited by tidal interaction (e.g. \cite{singh1996}).  The X-ray emission in active binaries systems comes from a late type companion which is spun up, driving a magnetic dynamo many times more powerful than the Sun's. The H$\alpha$ emission is produced in the active chromosphere and in prominences (also see \cite{pallavicini1981}). Our X-ray optical flux ratio for the star during quiecence (log(fx/fr) =-1.2) is in the upper end of the normal range for M stars \citep{zombeck}. 

The optical and X-ray observations described by \cite{mathu1990} for the By Dra star G 182 before, during and after a large flare in 1983 are similar to those found here (dM0.5e, B-V=1.4, V-R=0.9, V-I=1.85).  They found a quiescent X-ray luminosity of L$_X$=10$^{29.3}$ for G 182, and EW(H$\alpha \sim$2 \AA. Our measured EW(H$\alpha$) is however an order of magnitude larger. An optical spectrum will be interesting to determine whether CXO J005428.9-723107 shows other signs of enhanced activity, and/or extreme youth, such as prominent Lithium and Calcium lines.  

A search of SIMBAD database revealed a high proper motion star  OGLE 324153 \citep{soszynski2002}\footnote{CDS/Vizier catalog ID: J/AcA/52/143} at J2000 coords 00:54:28.66, -72:31:06.5 PM(RA)=158.1mas/yr, PM(Dec)=-0.1mas/yr V=19.486, B-V=1.626, V-I=3.017. Given the close match in photometry, the OGLE I-band reference image of field SMC-SC6, taken 11-08-1997 was compared with our Magellan I-band image taken 10 years later.  The counterpart is clearly OGLE 324153 and a motion of 1.6'' is apparent, consistent with that predicted by the OGLE proper motion (1.58''). The Magellan images were taken 15 months after the Chandra observation, so the motion during that time was +0.197'' in RA. This shift is smaller than the Chandra error circle, but if accounted for improves the alignment. For our preferred distance of 250 pc the projected tangential velocity is 200 km/s, such a velocity is typical for members of the Galactic halo. According to \cite{fehrenbach1984} the density of high-velocity foreground stars towards the Magellanic Clouds appears to be about 20 times higher than in other directions of similar galactic latitude. 

\subsection{Alternative Scenarios}
If the star lies in the SMC, with distance modulus of 18.5, then M$_V$ $\approx$+0.7, a red giant (M0III) would be required to match the photometry. So-called symbiotic stars (red giant with accreting White dwarf companion) can be bright X-ray sources.  Symbiotics display a great diversity of spectra, including thermal and non-thermal components, as can be seen in the ROSAT data presented 
by \cite{muerset1997}.  

Interpreting this object as a Be-NS binary in the SMC has several problems:  On the X-ray side we have the flare, a lack of pulsations and a softer spectrum than the ubiquitous $\Gamma$=1$\pm$0.5 powerlaw.  In the optical, the star is under-luminous for a Be star, by about 3 magnitudes, and in addition is too red in B-V.  According to \cite{mcbride2008}, of 37 spectrally confirmed BeX systems in the SMC all have spectral types earlier than B3; the mode is B0. From the observations listed in \cite{mcbride2008}, the faintest apparent V magnitude is 16.9, and the reddest color index is B-V=+0.15. In Be stars the B-V color is relatively unaffected by the red-excess, and is representative of the intrinsic spectral type. Thus the observed value of +1.4 rules out a Be star.  

The median M$_V$ for a CV secondary is +8 \citep{cvluminosityfn}, which at the SMC would be V= 26.5. During outbursts this rises by 5 magnitudes or more, and so a CV could become visible, but this is unlikely as the accretion disk emission would produce a blue B-V index. Similar brightness arguments apply to LMXBs: the star is too bright to be a CV or LMXB at the distance of the SMC,  If however it is a foreground galactic system, the M$_V$ constraint vanishes. 

\section{Conclusion}
We found that CXO J005428.9-723107 / OGLE 324153 is an active star or binary located relatively nearby, in the thick disk or halo of the Galaxy. The time-scale of the X-ray flare, and the spectrum during and afterwards share key characteristics with flaring M-stars (see for example \cite{wargelin2008} \& \cite{scholz2005}). The optical magnitude and color indices of the counterpart are consistent with an M0-M5 depending on the metallicity, amount of circumstellar or chromospheric emission (indicated by the strong H$\alpha$ emission) and contribution from a companion if present. The balance of evidence points to an active M5-V at $\sim$ 250pc with flare luminosity L$_X$=4.27$\times$10$^{30}$.  The high space velocity (200 km/s) and apparently low metallicity together suggest the star is in the halo. A binary is then needed to explain its highly active nature. 
 
The discovery of a flare star in this field is a cautionary example that an X-ray source with emission line optical-counterpart toward the SMC is not necessarily an HMXB. The SMC Deep Fields X-ray catalog contains several other sources whose quantiles suggest they are also stellar corona. These objects are the subject of a wider study which includes a search for the most X-ray active stars {\em in} the SMC.

\section{Acknowledgements}
This work was supported by the Gemini Observatory, which is operated by the Association of Universities 
for Research in Astronomy, Inc., on behalf of the international Gemini partnership of Argentina, 
Australia, Brazil, Canada, Chile, the United Kingdom, and the United States of America.
JJD was funded by NASA contract NAS8-39073 to the {\it Chandra X-ray
Center} during the course of this research. We thank Maureen van den Berg for providing the Magellan images.

\newpage

\end{document}